\documentclass[aps,prl,twocolumn,superscriptaddress,amsmath,amssymb,showpacs]{revtex4}

\usepackage{graphicx}
\usepackage{dcolumn}
\usepackage{bm}

\begin{document}
\title{Coexistence of the spin-density-wave and superconductivity in the Ba$_{1-x}$K$_x$Fe$_2$As$_2$}

\author{H. Chen}
\affiliation{Hefei National Laboratory for Physical Science at Microscale 
and Department of Physics, University of Science and Technology of China, 
Hefei, Anhui 230026, China}
\author{Y. Ren}
\affiliation{Advanced Photon Source, Argonne National Laboratory, Argonne, IL 60439, USA}
\author{Y. Qiu}
\affiliation{NIST Center for Neutron Research, National Institute of Standards
and Technology, Gaithersburg, MD 20899, USA}
\affiliation{Department of Materials Science and Engineering, University of Maryland, College Park, MD 20742, USA}
\author{Wei Bao}
\email{wbao@lanl.gov}
\affiliation{Los Alamos National Laboratory, Los Alamos, NM 87545, USA}
\author{R. H. Liu}
\author{G. Wu}
\author{T. Wu}
\author{Y. L. Xie}
\author{X. F. Wang}
\affiliation{Hefei National Laboratory for Physical Science at Microscale 
and Department of Physics, University of Science and Technology of China, 
Hefei, Anhui 230026, China}
\author{Q. Huang}
\affiliation{NIST Center for Neutron Research, National Institute of Standards
and Technology, Gaithersburg, MD 20899, USA}
\author{X. H. Chen}
\email{chenxh@ustc.edu.cn}
\affiliation{Hefei National Laboratory for Physical Science at Microscale 
and Department of Physics, University of Science and Technology of China, 
Hefei, Anhui 230026, China}

\date{\today}

\begin{abstract}
The relation between the spin-density-wave (SDW) and superconducting order is a central topic in current research on the FeAs-based high $T_C$ superconductors. Conflicting results exist
in the LaFeAs(O,F)-class of materials, for which whether the SDW and superconductivity are mutually exclusive or they can coexist has not been settled. Here we show that for the 
(Ba,K)Fe$_2$As$_2$ system, the SDW and superconductivity can coexist in an extended range of compositions. The availability of single crystalline samples and high value of the energy gaps would make the materials a model system to investigate the high $T_C$ ferropnictide superconductivity.

\end{abstract}


\maketitle

The FeAs-based new superconductors have generated great excitement,
since the superconducting transition temperature $T_C$ of these single-layer materials  surpasses that of the single-layer high $T_C$ cuprate superconductors\cite{A033603,A033790,A034234,A042053,A042105,A042582,A044290}.
In addition to the superconductivity\cite{Kamihara2008,A054630}, a closely related electron-pair instability on the Fermi surface, the spin-density-wave (SDW)\cite{sdw_awo}, has been inferred from the nesting Fermi surface\cite{A033426} and observed directly in neutron diffraction experiments in both LaFeAsO (1111)\cite{A040795} and BaFe$_2$As$_2$ (122)\cite{A062776} classes of materials. It is difficult to imagine that the same electronic state would participate in both the superconducting and SDW electron-pairs instability of the Fermi surface,
and indeed the electron and hole-like Fermi sheets in (Ba,K)Fe$_2$As$_2$ have been observed in angle-resolved photoemission spectroscopy (ARPES) study to be gapped by either the SDW\cite{A062627} or superconducting\cite{A070398,A070419} order in the parent or superconducting ($x=0.4$) compound respectively. Thus, it would be extremely informative to investigate the relationship between the two correlated electronic states in the phase diagram.

It should be noted that the pronounced anomaly in resistivity which was initially thought as indicating the SDW transition in LaFeAsO\cite{A033426} is actually associated with a tetragonal to orthorhombic structural transition, and the SDW occurs in another separate phase transition at a lower temperature\cite{A040795}. In BaFe$_2$As$_2$, the situation is simpler with the structural and SDW transitions concur at the same temperature $T_S$\cite{A062776}. So far, two types of phase diagrams have been reported for the 1111-class of materials. In the first type of phase diagram, before the orthorhombic phase of the parent compound is completely suppressed by doping, superconducting phase has
already appeared at low temperature as in LaFeAs(O,F)\cite{Kamihara2008} and SmFeAs(O,F)\cite{A042105,A063533}. Namely, there is an overlapping composition region where the SDW and superconductivity may coexist, although direct measurements of the SDW have not been conducted. For the second type of phase diagram, superconductivity appears after the SDW order is completely suppressed as reported for CeFeAs(O,F)\cite{A062528} and LaFeAs(O,F)\cite{A063533}. The conflicting reports in [\onlinecite{Kamihara2008}] and [\onlinecite{A063533}] for the LaFeAs(O,F) system have not yet been resolved.

In the ground-breaking work on Ba$_{1-x}$K$_x$Fe$_2$As$_2$ by
Rotter et al., only $x=0$, 0.4 and 1 samples were investigated\cite{A054630}.
For Ca$_{1-x}$Na$_x$Fe$_2$As$_2$, only the parent and one superconducting sample $x=0.5$ were studied\cite{A064279}.
The Sr$_{1-x}$K$_x$Fe$_2$As$_2$ system has been studied by Chen et al.\ for
$0\le x\le 0.4$ \cite{A061209} and by Sasmal et al.\ in the whole composition range for $0\le x\le 1$ \cite{A061301}, respectively. However, the focus was on the superconducting transition.
Magnetic and structural transitions were not investigated in these works on the Sr-122 materials. Here we report a systematic study on the Ba$_{1-x}$K$_x$Fe$_2$As$_2$
($0\le x\le 1$) system using transport, x-ray and neutron diffraction techniques. We establish that the Ba-122 system has the same type of phase diagram as the first type in the 1111-class materials. The structural and SDW transitions remains concurring at $T_S$ in the potassium doped samples until they both are suppressed at $x\sim 0.4$. Superconductivity appears for $0.2\le x\le 1$ in both the orthorhombic and tetragonal phases. Thus, for $0.2\le x< 0.4$, both the SDW and the superconducting orders coexist in the Ba$_{1-x}$K$_x$Fe$_2$As$_2$ system.

\begin{figure}
\includegraphics[width=.8\columnwidth,angle=0,clip=true]{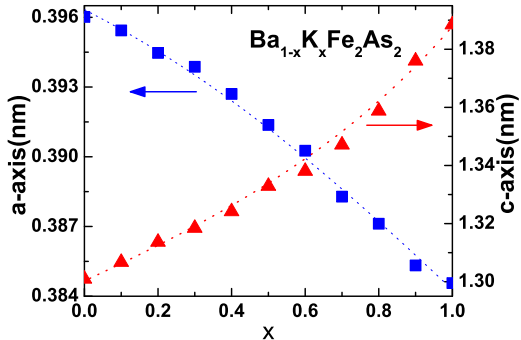}
\caption{The lattice parameters at room temperature as a function of the composition.}
\label{fig1}
\end{figure}

A series of polycrystalline Ba$_{1-x}$K$_x$Fe$_2$As$_2$ samples covering from the Ba ($x
= 0$) to the K ($x=1$) end member were synthesized by solid state reaction method
using BaAs, KAs and Fe$_2$As as the starting materials. The BaAs was
pre-synthesized by reacting Ba powder with As powder in an evacuated
quartz tube at 673 K for 4 hours; KAs by reacting K
lumps with As powder at 523 K for 4 hours; and Fe$_2$As by
reacting the mixture of element powders at 973 K for 4 hours. The
raw materials were weighed according to the
stoichiometric ratio of Ba$_{1-x}$K$_x$Fe$_2$As$_2$. The
mixed powders were then thoroughly grounded and pressed into pellets.
The pellets were wrapped with Ta foil and sealed in evacuated quartz
tubes. These tubes were annealed at $973 \sim 1093$ K for 20 hours.
The sample preparation, except for the annealing, was carried out
in glove box in which high purity argon atmosphere is filled. The samples
were characterized using an X-ray diffractometer at room temperature and the
powder diffraction pattern of 
all samples can be indexed using the tetragonal ThCr$_2$Si$_2$ structure of the space group $I4/mmm$ (No~139). The variation of the lattice parameters at room temperature is shown in Fig.~\ref{fig1}. Both the $a$ and $c$ change smoothly from $x=0$ to 1. 

\begin{figure}
\includegraphics[width=.9\columnwidth,angle=0,clip=true]{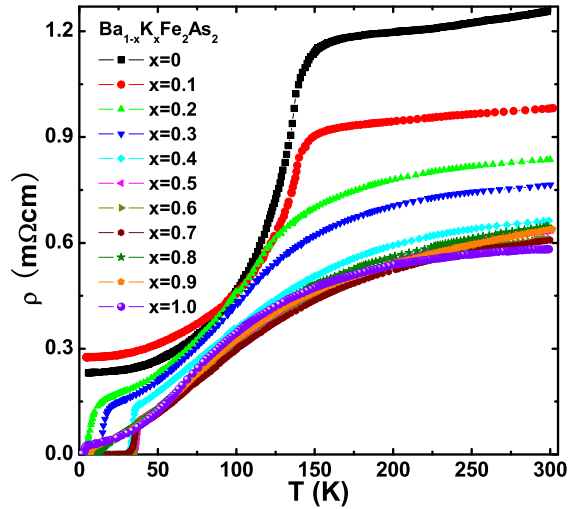}
\caption{Temperature dependence of the resistivity. 
}
\label{fig2}
\end{figure}

\begin{figure}
\includegraphics[width=.9\columnwidth,angle=0,clip=true]{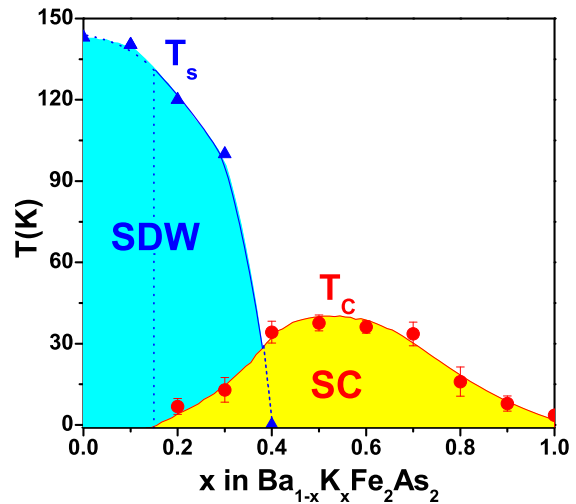}
\caption{The composition-temperature phase-diagram, showing the structural, magnetic and superconducting transitions. The $T_S$ denotes the temperature of the simultaneous structural and magnetic transition, and $T_C$ the superconducting one. The spin-density-wave (SDW) and superconducting (SC) orders coexist at low temperature in $0.2\le x< 0.4$.}
\label{fig3}
\end{figure}

We measured the resistivity using the standard four-probe method. The results are shown in Fig.~\ref{fig2}.
The anomaly associated with the structural and magnetic transition is pronounced for $x=0$ and 0.1. The anomaly is rounded off for $x=0.2$,
which becomes a superconductor with the transition starting at 14 K and
the resistivity reaching zero at 3 K. The $T_C$ increases with further potassium doping and the superconducting transition becomes narrower until $x=0.5$. Thereafter, $T_C$ begin to decrease from the maximum $T_C\approx 37.5$ K with the potassium doping. At $x=1$, $T_C$ is 3.8 K for the
KFe$_2$As$_2$ sample, the same as reported by Sasmal et al.\cite{A061301}. The $T_C$ as a function of the composition is summarized in Fig.~\ref{fig3}.

\begin{figure}
\includegraphics[width=.7\columnwidth,angle=90,clip=false]{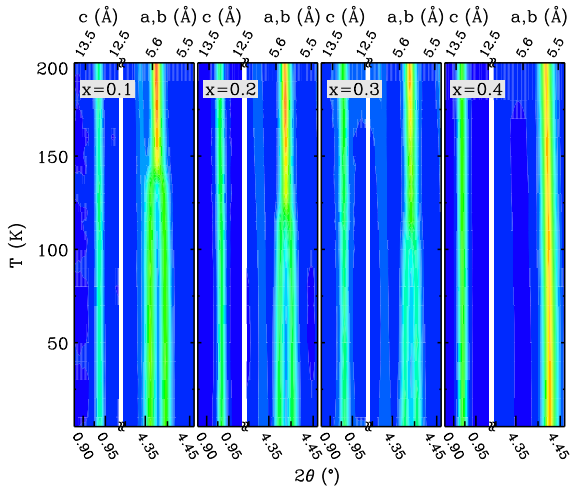}
\caption{The temperature dependence of the structural Bragg peak (002) near $2\theta=0.95^o$ and (220) near $4.38^o$ for $x=0.1$, 0.2, 0.3 and 0.4. The splitting of the (220) peak, in the tetragonal notation, indicates the tetragonal to orthorhombic structural transition.}
\label{fig4}
\end{figure}

While the temperature of the simultaneous structural and SDW transition can be inferred from the pronounced anomaly in resistivity for $x=0$ and 0.1  (Fig.~\ref{fig2}), for $x\ge 0.2$, it becomes progressively less certain whether there is an anomaly in the resistivity.
To further investigate the crystal structure and structural transition, powder diffraction experiments from 5 to 300 K were performed for the $x=0$, 0.1, 0.2, 0.3, 0.4 and 0.6 samples using synchrotron high-energy x-ray ($\lambda=0.1067 \AA$) at the beamline 11-ID-C at Advanced Photon Source of ANL.
The sample temperature was controlled by a cryomagnet. The synchrotron X-ray Bragg peaks from all samples are resolution-limited, indicating uniform sample quality and excluding the possibility of phase separation. The full spectra and detailed structural study will be reported elsewhere. Here in 
Fig.~\ref{fig4}, we shows the temperature dependence of the (002) and (220) Bragg reflections,
labeled using the high-temperature tetragonal $I4/mmm$ unit cell, for $x=0.1$-0.4 in the 5 to 200 K temperature interval. 
The 
(002) peak at $2\theta\sim 0.95^o$ traces the temperature dependence of the $c$ lattice
parameter. The (220) peak at $2\theta\sim 4.38^o$ is useful for tracing the in-plane lattice parameters and its splitting signifies the orthorhombic structural transition\cite{A054630,A062776}. It is obvious that the structural transition occurs for the $x=0.1$-0.3
samples, while the transition is absent for the $x=0.4$ sample down to 5 K.

\begin{figure}
\includegraphics[width=.7\columnwidth,angle=90,clip=true]{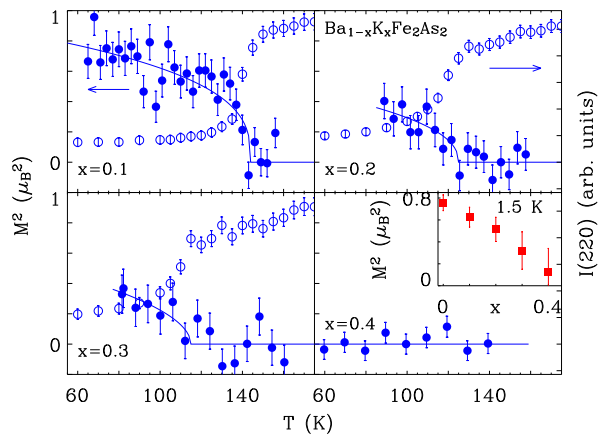}
\vskip -.3cm
\caption{The squared magnetic order parameter measured with neutron diffraction at magnetic (101) Bragg reflection of the orthorhombic phase is compared with the structural (220) Bragg peak of the high-temperature tetragonal phase. The squared magnetic moment at 1.5 K is shown as a function of doping in the inset. The orthorhombic distortion and the SDW order concur below $x<0.4$ at low temperature.}
\label{fig5}
\end{figure}

In Fig.~\ref{fig5}, the X-ray diffraction intensity at the fixed $2\theta$ position of the (220) peak at high temperature is shown as a function of the temperature for $x=0.1$-0.3 (open circles).
Albeit less abrupt than in BaFe$_2$As$_2$\cite{A062776}, the intensity curve of the $x=0.1$ sample maintains the first-order drop at $T_S\approx 144$ K, corroborating the pronounced anomaly in resistivity (Fig.~\ref{fig2}) and the abrupt splitting of the (220) peaks in Fig.~\ref{fig4}. Consistent with the more gradual splitting of the (220) peaks and smoother variation of the resistivity for $x=0.2$ and 0.3, the intensity of the (220) in Fig.~\ref{fig5} reduces gradually below $T_S\approx 125$ and 115 K, respectively, similar to the second-order structural transition of LaFeAsO\cite{A040795}.

Magnetic Bragg peaks due to the SDW antiferromagnetic order\cite{A062776} were measured using the triple-axis neutron spectrometer BT9 at the NIST Center for Neutron Research. The neutron
beam of 14.7 meV was selected using the (002) reflection of pyrolytic graphite as both the monochromator and analyzer. Pyrolytic graphite filters of 10 cm total thickness were placed in the neutron beam path to reduce higher order neutrons. The sample temperature was controlled by a pumped He cryostat. The squared staggered magnetic moment determined from the magnetic (101) and structural (002) Bragg peaks measured at 1.5 K is shown in the inset to Fig.~\ref{fig5}. The potassium doping smoothly suppresses the magnetic moment $M$, which approaches zero at $x\approx 0.4$.
To verify the relation between the structural and the magnetic transitions, we measured the magnetic (101) Bragg intensity as a function of temperature (solid circles in Fig.~\ref{fig5}). The N\'{e}el temperature is indistinguishable from the $T_S$. Thus, the magnetic and structural transitions concur in the Ba$_{1-x}$K$_x$Fe$_2$As$_2$ system.

The $T_S$ as a function of doping is shown in the phase-diagram in Fig.~\ref{fig3}. Below $T_S$, the SDW order exists on the orthorhombically distorted lattice. There are three
composition regions for the Ba$_{1-x}$K$_x$Fe$_2$As$_2$ system. I) Near the BaFe$_2$As$_2$ end for small $x<0.2$, the first-order structural transition is associated with a pronounced anomaly
in resistivity. There is no superconductivity in this composition region and electrons at the Fermi surface form the SDW order below $T_S$. II) At the other end of the phase-diagram for $x>0.4$, the SDW order disappears and there exists only superconductivity below $T_C$. Above $T_C$, resistivity is comparable in the composition region (Fig.~\ref{fig2}). III)
Between the two composition regions, the structural transition becomes second-order, the resistivity anomaly become smeared out, and both the SDW and superconductivity coexist at the low temperature.

One trivial possibility for the coexistence of the SDW and superconductivity is the phase separation, namely the $x=0.25$ sample is a mixture of, e.g., the $x=0.1$ and 0.4 samples.
However, the resolution of our synchrotron X-ray experiments clearly rules this possibility out. Thus the coexistence is an intrinsic property of the Ba$_{1-x}$K$_x$Fe$_2$As$_2$ system.

Multiple electronic bands cross the Fermi level in Ba$_{1-x}$K$_x$Fe$_2$As$_2$, leading to multiple Fermi sheets. Four Fermi sheets have been observed directly in
the ARPES experiments in the composition region I\cite{A062627,A063453} and three Fermi sheets observed in the region II\cite{A070398,A070419}. The phase-diagram in Fig.~\ref{fig3}resembles that for another multiple-Fermi-sheet heavy-fermion superconductor system CeRh$_{1-x}$Ir$_x$In$_5$\cite{bao04b}. The high $T_C$ of the Ba-122 system, however, provides opportunity to directly measure the gaps caused by electron-pair order on the Fermi sheets using available technique of the ARPES. For BaFe$_2$As$_2$ in the region I, one Fermi sheet near the M point and probably another one near the $\Gamma$ point are gapped by the SDW order\cite{A062627}. For Ba$_{0.6}$K$_{0.4}$Fe$_2$As$_2$ in the region II, all three Fermi sheets are gapped by the superconducting order without nodes\cite{A070398,A070419}, consistent with the Andreev reflection measurements\cite{A054616}. The sole Fermi sheet near the M point and the inner Fermi sheet near the $\Gamma$ point have a larger superconducting gap than that on the outer Fermi sheet near $\Gamma$ point. Now we have found that both the SDW and superconductivity coexist in the region III. It would be extremely interesting to investigate sample in the composition region to directly observe the gaps on the Fermi sheets when the sample goes through successive SDW and superconducting transitions.

In summary, we have synthesized high quality samples of the Ba$_{1-x}$K$_x$Fe$_2$As$_2$ in the whole composition range. Combining transport, X-ray and neutron diffraction studies, we demonstrate that the SDW order from the Ba end and the superconductivity from the K end overlap in the intermediate composition range. The availability of single crystalline samples and the high $T_C$/$T_S$ thus large energy gaps would allow current experimental techniques to directly investigate the interplay between the itinerant magnetism and superconductivity in microscopic details in this 122 system. Therefore, the system investigated in this work is likely to tremendously enhance our understanding on high-$T_C$ superconductivity in the FeAs-based superconductors in particular and magnetic superconductors in general.

The work at USTC is supported by the
Nature Science Foundation of China and by the Ministry of Science
and Technology of China (973 project No: 2006CB601001) and by
National Basic Research Program of China (2006CB922005); at LANL by U.S.\ DOE/OS/BES; at ANL by U. S. DOE/OS/BES, under Contract No. DE-AC02-06CH11357.

\end{document}